\documentclass[aps,prl,twocolumn,showpacs,superscriptaddress]{revtex4-1}
\usepackage{graphicx}
\usepackage{dcolumn}
\usepackage{bm}
\usepackage{amssymb}
\usepackage{amsmath}
\usepackage{subfigure}
\usepackage{float}
\begin{document}
\title{Dynamics of Order Parameter in Photoexcited Peierls Chain}
\author{Yong Wang}
\affiliation{Department of Physics, The University of Hong Kong, Hong Kong SAR, China}
\author{Wei-Qiang Chen}
\affiliation{Department of Physics, South University of Science and Technology of China, China}
\author{Fu-Chun Zhang}
\affiliation{Department of Physics, Zhejiang University, China}
\affiliation{Department of Physics, The University of Hong Kong, Hong Kong SAR, China}
\begin{abstract}
The photoexcited dynamics of order parameter in Peierls chain is investigated by using a microscopic quantum theory in the limit where the hot electrons may establish themselves into a quasi-equilibrium state described by an effective temperature. The optical phonon mode responsible for the Peierls instability is coupled to the electron subsystem, and its dynamic equation is derived in terms of the density matrix technique. Recovery dynamics of the order parameter is obtained, which reveals a number of interesting features including  the change of oscillation frequency and amplitude at phase transition temperature and the photo-induced switching of order parameter.
\end{abstract}
\maketitle
\emph{Introduction}
The concept of order parameter, which characterizes various phases in condensed matters, plays an essential role in modern phase transition theory\cite{OrdP}. One of the recent focuses is the dynamics of order parameter in non-equilibrium process, which is of interest in both basic and applied physics.  Micromagnetics for magnetoelectronics devices is an excellent example of such applications\cite{MicroM}. The non-equilibrium dynamics of the charge density wave(CDW) order has recently been extensively investigated by using optical pump-probe spectroscopy\cite{probe1,probe2,probe3,probe4} and time-dependent angle-resolved photoemission spectroscopy (trARPES)\cite{ARPES1,ARPES2,ARPES3,ARPES4,ARPES5,ARPES6}. In these experiments, the samples are initially excited by a pump laser pulse, and the electronic dynamics is then observed either by probing the laser pulse or by using the ARPES technique. Depending upon the pump fluence, there are two types of electronic dynamics. In the low pump fluence case, the CDW order is only slightly perturbed and the collective modes in the CDW phase are excited. If the pump fluence is high enough to destroy(meltdown) the CDW order, a dynamic phase transition occurs and the recovery dynamics of the initial CDW order will be observed. These two types of dynamics are not mutually exclusive but closely related, since the recovery of the CDW order is always accompanied with the excitations of the collective mode\cite{probe1,probe2,probe3,probe4,ARPES1,ARPES2,ARPES3,ARPES4,ARPES5,ARPES6}. The study of these nonequilibrium dynamics processes may provide unique insight to identify the formation mechanism of the CDW order\cite{ARPES1,ARPES4,ARPES5}. In comparison with the experimental progresses, the theoretical efforts to understand and describe these phenomena are still very limited. One approach is the time-dependent Ginzburg-Landau (TDGL) equation\cite{probe4}. As a phenomenological description, though it is effective in describing some experiments, it does not give the microscopic details of the dynamics. Considering the complex microscopic interactions in these CDW materials, it is helpful and desirable to develop a microscopic theory for more fundamental and complete understanding of the dynamic processes. Such a necessity has been exemplified in the study of the optical conductivity of the Hubbard-Holstein model\cite{HHM}. In this Letter, we consider the photoexcited dynamics of the Peierls-type CDW systems based on a microscopic model. We consider half-filled Peierls chain as a minimal model, and show how the nonequilibrium dynamics of the order parameter emerges from the microscopic interaction and how the coefficients in the TDGL equation are related to the basic parameters in the microscopic model.
\begin{figure}[htbp]
\includegraphics[scale=0.25,clip]{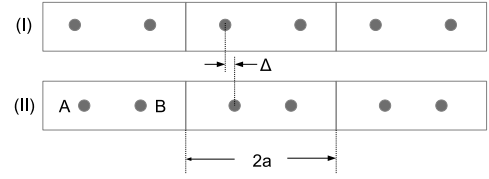}
\caption{(Color online)(I) Undistorted and (II) distorted configurations of a Peierls chain. The chain consists of $N$ unit cells with two atoms ($\eta=A,B$) in each unit cell. The lattice constant is $2a$, and the displacement of the distorted $A$-type atoms is the order parameter $\Delta$.}\label{PC}
\end{figure}

\emph{Theoretical Model} We consider a Peierls chain consisting of $2N$ monovalent atoms with the periodic boundary condition as depicted in Fig.~\ref{PC}.  The equilibrium positions of the atoms at high temperature are shown in Configuration (I). The deviation of the $i$th atom from its equilibrium position is denoted as $u_{i}$. The system Hamiltonian, or the Su-Schrieffer-Heeger model, is defined as\cite{SSH}
\begin{eqnarray}
H&=&-\sum_{i,\sigma}(t_{i,i+1}c_{i+1,\sigma}^{\dag}c_{i,\sigma}+h.c.)\nonumber\\
&+&\sum_{i}\frac{1}{2}M\dot{u}_{i}^{2}+\sum_{i}\frac{1}{2}K(u_{i+1}-u_{i})^{2},\label{H0}
\end{eqnarray}
where we only consider the nearest neighbor hopping of the electrons.  $c_{i,\sigma}^{\dag}$($c_{i,\sigma}$) is the creation (annihilation) operator for the electron at $i$th atom with spin $\sigma$, $M$ is the atom mass, and $K$ characterizes the force constant to pull the atom back to its equilibrium position.  The hopping integral $t_{i, i+1}$ is function of $\delta u_i = u_i - u_{i+1}$.  Since $\delta u_{i}$ is usually small, we can expand $t_{i, i+1}$ to the first order of $\delta u_i$, i.e.  $t_{i+1,i}=t_{0}-\alpha(u_{i+1}-u_{i})$, where $t_0$ is the hopping integral in the absence of
lattice distortion. 

At low temperatures, the half-filled chain will be dimerized as shown in configuration (II) due to the Peierls
instability.  The unit cell is enlarged to include two atoms in this phase, so the lattice constant becomes $2a$ with $a$ the nearest neighbor atom distance in the un-dimerized state. In the
following, we use the enlarged unit cell for convenience, where the unit cell is denoted with $m=1,2,...,N$, and the
two atoms in one unit cell are denoted as $A$ and $B$ respectively.  Then we move to the reciprocal in
the reduced Brillouin zone space\cite{supp}, and the Hamiltonian reads
\begin{eqnarray}
H_{p}&=&\frac{1}{2}\sum_{\mu,q}(\widehat{P}_{\mu,q}^{*}\widehat{P}_{\mu,q}+\omega_{\mu,q}^{2}\widehat{Q}_{\mu,q}^{*}\widehat{Q}_{\mu,q}),\label{H2p}\\
H_{e}&=&-t_{0}\sum_{k,\sigma}[(1+e^{-2ika})c_{A,k,\sigma}^{\dag}c_{B,k,\sigma}+h.c.],\label{H2e}\\
H_{ep}&=&\sum_{\mu,q,k,\sigma}(g_{\mu,q,k}\widehat{Q}_{\mu,q}c_{A,k,\sigma}^{\dag}c_{B,k-q,\sigma}+h.c.),\label{H2ep}
\end{eqnarray}
where $H_p$ and $H_{e}$ are the Hamiltonian of phonon and electron subsystem respectively, $H_{ep}$ describes the electron-phonon interaction, $q,k\in [-\frac{\pi}{2a},\frac{\pi}{2a}]$ are the wavevectors in the reduced zone. $\mu=a,o$ denotes the acoustic and optical phonon respectively.  $\widehat{Q}_{\mu,q}$ and $\widehat{P}_{\mu,q}$ are the canonical coordinates and momentums of the phonons respectively, and they satisfy the quantization condition $[\widehat{Q}_{\mu,q},\widehat{P}_{\mu',q'}]=i\hbar\delta_{q,-q'}\delta_{\mu,\mu'}$. $\omega_{\mu,q}$ is the
phonon energy. $c_{\eta,k,\sigma}^{\dag}$($c_{\eta,k,\sigma}$) is the Fourier transformation of
$c_{\eta,m,\sigma}^{\dag}$($c_{\eta,m,\sigma}$), with $\eta=A,B$. $g_{\mu,q,k}$ represents the spin-independent electron-phonon
interaction.

For the half-filled Peierls chain, the nesting wavevector is $\mathbf{K}_{nest} = 0$ in the reduced zone. So the phonon mode which condenses in the Peierls phase transition is the optical one with $q=0$, denoted as $(o,0)$.  In the pump-probe technique, the electron subsystem is excited to a highly non-equilibrium state by the strong pump laser. Then it reaches quasi-equilibrium state through electron-electron scattering in femtosecond time scale. The quasi-equilibrium state could be characterized by an effective temperature $T_{e}$.  On the other hand, the lattice is nearly unaffected in the thermalization process because of the much larger mass of the ions.  It is reasonable to assume
that the lattice keeps its original temperature $T_l$. In the Peierls phase, the distortion of the lattice is due to the electron-phonon coupling.  In the case where only the electrons are heated, it is obviously that the ions are not at the configuration with lowest energy.  So the ions will start to move towards the configuration with lower energy.  On the other hand, the eigenstates of the electrons also varies with the movement of the ions.  Meanwhile, the hot electrons will transfer the extra energy to the lattice, which could be considered as a heat bath with temperature $T_l$, until its temperature $T_{e}$ reduced to $T_{l}$.  Based on this picture, we use two equations to describe the dynamics of the photoexcited Peierls chain. Since we are more interested in the dynamics of the lattice, we will first derive the equation describing the dynamics of the condensate of phonon mode $(o,0)$, with its coupling to the electrons in the microscopic Hamiltonian through density matrix technique. Then we introduce a phenomenological equation to describe the temperature decay of the electron subsystem.

In the following, we denote the operators $(\widehat{Q}_{o,0},\widehat{P}_{o,0})$ for the phonon mode
$(o,0)$ as $(\widehat{\mathsf{Q}},\widehat{\mathsf{P}})$, and the corresponding frequency as
$\Omega^{2}\equiv\omega_{o,0}^{2}=\frac{4K}{M}$. With the density matrix $\rho$ of the phonon mode $(o,0)$, the expectation value of the operators are $\mathcal{Q}(t)=\text{Tr}[\widehat{\mathsf{Q}}\rho]$,$\mathcal{P}(t)=\text{Tr}[\widehat{\mathsf{P}}\rho]$.  To investigate the dynamics of $\mathcal{Q}$, we will study how it evolves from time $t'$ to $t = t' + \delta t$ within a small $\delta t$.  For simplicity, we denote $\mathcal{Q}_0\equiv\mathcal{Q}(t')$, and we define a new canonical coordinate $\widehat{\mathsf{q}}$ as $\widehat{\mathsf{q}}\equiv\widehat{\mathsf{Q}}-\mathcal{Q}_0$. Then the Hamiltonian of the phonon mode $(o,0)$ becomes
\begin{eqnarray}
H_{S}=\frac{1}{2}[\widehat{\mathsf{P}}^{2}+\Omega^{2}(\widehat{\mathsf{q}}^{2}+2\mathcal{Q}_{0}\widehat{\mathsf{q}}+\mathcal{Q}_{0}^{2})].\label{HS}
\end{eqnarray}
The coupling term for the electrons and the phonon mode $(o,0)$ contained in Eq.~(\ref{H2ep}) is also decomposed into two terms $-\mathcal{Q}_0 \widehat{\mathsf{F}}$ and $-\widehat{\mathsf{q}}\otimes\widehat{\mathsf{F}}$, where
\begin{eqnarray}
\widehat{\mathsf{F}}=\sqrt{\frac{2\alpha^{2}}{NM}}\sum_{k,\sigma}[(1-e^{-2ika})c_{A,k,\sigma}^{\dag}c_{B,k,\sigma}+h.c.].\label{Force}
\end{eqnarray}
In the small time interval $[t',t]$, the effect of the Peierls distortion on the hot electrons is dominated by the term $-\mathcal{Q}_0 \widehat{\mathsf{F}}$, which gives the electron Hamiltonian as $H_{B}=H_{e}-\mathcal{Q}_{0}\widehat{\mathsf{F}}$. On the other hand, the term $V\equiv -\widehat{\mathsf{q}}\otimes\widehat{\mathsf{F}}$ gives the quantum dynamics of the phonon mode $(o,0)$ driven by the hot electrons.

We now focus on the quantum dynamics of the phonon mode $(o,0)$ during $[t',t]$.  In the two temperature scenario, we assume that the quasi-equilibrium state of the hot   electrons has been established with the effective temperature $T_e$.  Then the phonon dynamics can be calculated with the standard ``system plus reservoir" paradigm, where the system Hamiltonian is
$H_{S}$, the electron reservoir Hamiltonian is $H_{B}$, while the interaction is $V$.  To the second order of the interaction $V$, the equation of motion of the density matrix of the phonon mode $(o,0)$ in the interaction picture is\cite{supp,denm}
\begin{eqnarray}
\frac{d\widetilde{\rho}(t)}{dt}&=&-\frac{1}{i\hbar}\langle\widetilde{\mathsf{F}}\rangle_{B}[\widetilde{\mathsf{q}}(t),\widetilde{\rho}(t')]\nonumber\\
&-&\frac{1}{\hbar^{2}}\int_{t'}^{t}d\tau\mathcal{J}(t-\tau)[\widetilde{\mathsf{q}}(t),[\widetilde{\mathsf{q}}(\tau),\widetilde{\rho}(\tau)]]\nonumber\\
&-&\frac{i}{\hbar^{2}}\int_{t'}^{t}d\tau\mathcal{K}(t-\tau)[\widetilde{\mathsf{q}}(t),\{\widetilde{\mathsf{q}}(\tau),\widetilde{\rho}(\tau)\}].\label{den}
\end{eqnarray}
Here, $\widetilde{\ldots}$ denotes the operators in the interaction picture, and
$\langle\ldots\rangle_{B}\equiv\text{Tr}_{B}[\ldots\widetilde{\rho}_{B}]$ denotes the expectation value over the reservoir density matrix $\widetilde{\rho}_{B}$. $\mathcal{J}(t-\tau)$ and $\mathcal{K}(t-\tau)$ are the real and imaginary part of the correlation function $\langle\widetilde{\mathsf{F}}(t)\widetilde{\mathsf{F}}(\tau)\rangle_{B}$ respectively. Based on physical considerations, the above equation holds if $\delta t$ is larger than the correlation time of the kernels $\mathcal{J}(t-\tau)$ and $\mathcal{K}(t-\tau)$ but still small enough in order to assume that $\mathcal{Q}$ and $T_{e}$ are unchanged during the time interval. This can be justified since the lattice dynamics is on the picosecond time scale while the kernel correlation time is in the order of femtosecond.

With Eq.~(\ref{den}) for the density matrix, the equations for $\dot{\mathcal{Q}}$ and $\dot{\mathcal{P}}$ during the
small time interval $[t',t]$ are\cite{supp}
\begin{eqnarray}
\dot{\mathcal{Q}}(t)&=&\mathcal{P}(t),\label{dQdt}\\
\dot{\mathcal{P}}(t)&=&-\Omega^{2}\mathcal{Q}(t)+\mathcal{F}(t)-\frac{2}{\hbar}\int_{t'}^{t}d\tau\mathcal{K}(t-\tau)\mathsf{q}(\tau).\label{dPdt}
\end{eqnarray}
Here, $\mathcal{F}=\langle\widetilde{\mathsf{F}}\rangle_{B}$. The displacement of the ions in real space is $\Delta\equiv u_{A,m}=-u_{B,m}=\frac{\mathcal{Q}}{\sqrt{2NM}}$. By applying the partial integration to the convolution integral in Eq.~(\ref{dPdt}) and dropping the term proportional to $\mathsf{q}(t)$ which vanishes for $\delta t\rightarrow 0$, we get an equation for the order parameter $\Delta$\cite{supp}
\begin{eqnarray}
\ddot{\Delta}+\int_{t'}^{t}d\tau\gamma(t-\tau)\dot{\Delta}(\tau)-\frac{\mathsf{f}}{M}=0.\label{equD}
\end{eqnarray}
Here, the damping kernel $\gamma(t)$ is related to the kernel $\mathcal{K}(t)$ via their Fourier transformations $\gamma(\omega)=-\frac{2}{\hbar}\frac{\mathcal{K}(\omega)}{\omega}$, and $\mathsf{f}$ is the force conjugated to the order parameter. Eq.~(\ref{equD}) takes the similar form as the TDGL equation, but its validity is not restricted to the temperature range near the phase transition. Furthermore, the damping kernel $\gamma$ and the force $\mathsf{f}$ are not given phenomenologically but are determined from the parameters in the Peierls model including $K,M,t_{0},\alpha$ and $T_{e}$. This enables us to understand microscopic origin of the order parameter dynamics directly.

The force $\mathsf{f}$ in Eq.~(\ref{equD}) consists of two terms\cite{supp},
\begin{eqnarray}
\mathsf{f}&=&-4K\Delta+\frac{1}{N}\sum_{k}\frac{16\alpha^{2}\Delta\sin^{2}ka}{\Lambda_{k}}\text{tanh}\frac{\beta\Lambda_{k}}{2},\label{FOCE}
\end{eqnarray}
where $\beta=1/k_{B}T_{e}$.  The first term is the elastic force between the ions, and the second term comes from the
coupling to the electron subsystem. The spin index has been taken into account by a factor 2 in the summation.  The
dispersions of the electron bands are $\epsilon_{\pm,k,\sigma}=\pm\Lambda_{k}$ with
$\Lambda_{k}=\sqrt{4t_{0}^{2}\cos^{2}ka+16\alpha^{2}\Delta^{2}\sin^{2}ka}$. It is easy to check that the force
$\mathsf{f}$ is related to the free energy $\mathcal{E}$ as $\mathsf{f}=-\partial\mathcal{E}/\partial\Delta$, with
\begin{eqnarray}
\mathcal{E}(\Delta,T_{e})=2K\Delta^{2}-\frac{1}{N\beta}\sum_{k}\text{ln}(2+e^{-\beta\Lambda_{k}}+e^{\beta\Lambda_{k}}).\label{FEden}
\end{eqnarray}

The Fourier transformation of the kernel functions $\mathcal{J}(\omega),\mathcal{K}(\omega),\gamma(\omega)$ can be calculated based on the microscopic model. Especially, we have\cite{supp}
\begin{eqnarray}
  \gamma(\omega)=\frac{2}{\hbar\omega}\sum_{k}|h_{k}|^{2}\text{tanh}\frac{\beta\Lambda_{k}}{2}[
  \delta(\omega-\frac{2\Lambda_{k}}{\hbar})-\delta(\omega+\frac{2\Lambda_{k}}{\hbar})],\nonumber
\end{eqnarray}
which is is an even function of $\omega$.  Here, the coefficients $|h_{k}|^{2}$ are \cite{supp}
\begin{eqnarray}
|h_{k}|^{2}&=&\frac{8\alpha^{2}}{NM}\frac{t_{0}^{2}\sin^{2}ka\cos^{2}ka}{t_{0}^{2}\cos^{2}ka+4\alpha^{2}\Delta^{2}\sin^{2}ka}.\label{hk2}
\end{eqnarray}
Then the damping kernel in Eq.~(\ref{equD}) becomes
\begin{eqnarray}
\gamma(t)=\int_{-\infty}^{+\infty}d\omega\gamma(\omega)\cos(\omega t).\label{gamt}
\end{eqnarray}
In a typical Peierls system, $t_{0}$ is in the order of eV, while $\alpha\Delta$ and $1/\beta$ are in the order of
meV. Thus $\gamma(\omega)$ is affected by the order parameter $\Delta$ and the temperature $T_{e}$ merely in the low
frequency range characterized by the energy $1/\beta$ and $\alpha\Delta$, and $\gamma(t)$ may be approximated to be
independent on $T_{e}$ and $\Delta$. Besides, the maximum of $\Lambda_{k}$ is $2t_{0}$ since $t_{0}\gg\alpha\Delta$,
then the cutoff frequency of $\gamma(\omega)$ will be $\omega_{c}=4t_{0}/\hbar$.  The correlation time of the damping
kernel $\gamma(t)$ is estimated as $t_{c}=\frac{2\pi}{\omega_{c}}=\frac{\pi\hbar}{2t_{0}}$, therefore $t_{c}$ is in the
order of femtosecond, which is much shorter than the picosecond time scale of the order parameter dynamics. This
suggests that the Markov approximation can be applied to the damping term in Eq.~(\ref{equD}) by replacing
$\dot{\Delta}(\tau)$ with $\dot{\Delta}(t)$. Then Eq.~(\ref{equD}) is reduced to
\begin{eqnarray}
\ddot{\Delta}+\Gamma\dot{\Delta}-\frac{\mathsf{f}}{M}=0,\label{equD1}
\end{eqnarray}
where $\Gamma=\int_{t'}^{t}d\tau\gamma(t-\tau)=\int_{0}^{\delta t}d\tau\gamma(\tau)$ with $\delta t>t_{c}$.

In order to get a qualitative estimation of $\Gamma$, we consider the case $\Delta=0$ and $\beta\Lambda_{k}\gg 1$, which
gives $|h_{k}|^{2}=\frac{8\alpha^{2}}{NM}\sin^{2}ka$ and $\Lambda_{k}=2t_{0}\cos ka$. $\gamma(\omega)$ is calculated as
$\gamma(\omega)=\frac{4\alpha^{2}}{\pi Mt_{0}}\frac{\sqrt{1-(\omega/\omega_{c})^{2}}}{|\omega|}$ for
$|\omega|\le\omega_{c}$. Then one gets $\gamma(\omega)\sim\frac{4\alpha^{2}}{\pi
  Mt_{0}\omega_{c}}\sim\frac{\hbar\alpha^{2}}{\pi Mt_{0}^{2}}$,
$\gamma(t)\sim\gamma(\omega)\omega_{c}\sim\frac{4\alpha^{2}}{\pi Mt_{0}}$ and
$\Gamma\sim\gamma(t)t_{c}\sim\frac{\hbar\alpha^{2}}{Mt_{0}^{2}}$. So we can estimate the damping coefficient $\Gamma$ in
Eq.~(\ref{equD1}) from the basic parameters $M,t_{0},\alpha$ except for a proportional factor.

On the other hand, all the phonon modes except $(o,0)$ form a thermal reservoir with temperature $T_{l}$.  The electron
subsystem will dissipate extra energy into the reservoir through electron-phonon coupling in the picosecond time
scale. The two-temperature scenario can be applied to describe this dissipation process since the thermalization process
of hot electrons happens in femtosecond time scale. For simplicity, we assume that the electronic temperature $T_{e}$
decays towards $T_{l}$ in the exponential form, i.e.
\begin{eqnarray}
T_{e}(t)=T_{l}+(T_{e}(0)-T_{l})e^{-t/\kappa},\label{Tdcay}
\end{eqnarray}
where the initial electronic temperature $T_{e}(0)$ depends on the fluence of the pump laser, and the decay time $\kappa$ is a fitting parameter. Eq.~(\ref{equD1}) and Eq.~(\ref{Tdcay}) are the two basic equations in our theory to describe the subpicosecond dynamics of the photoexcited Peierls chain.

\begin{figure}[htbp]
\includegraphics[width=0.23\textwidth,clip]{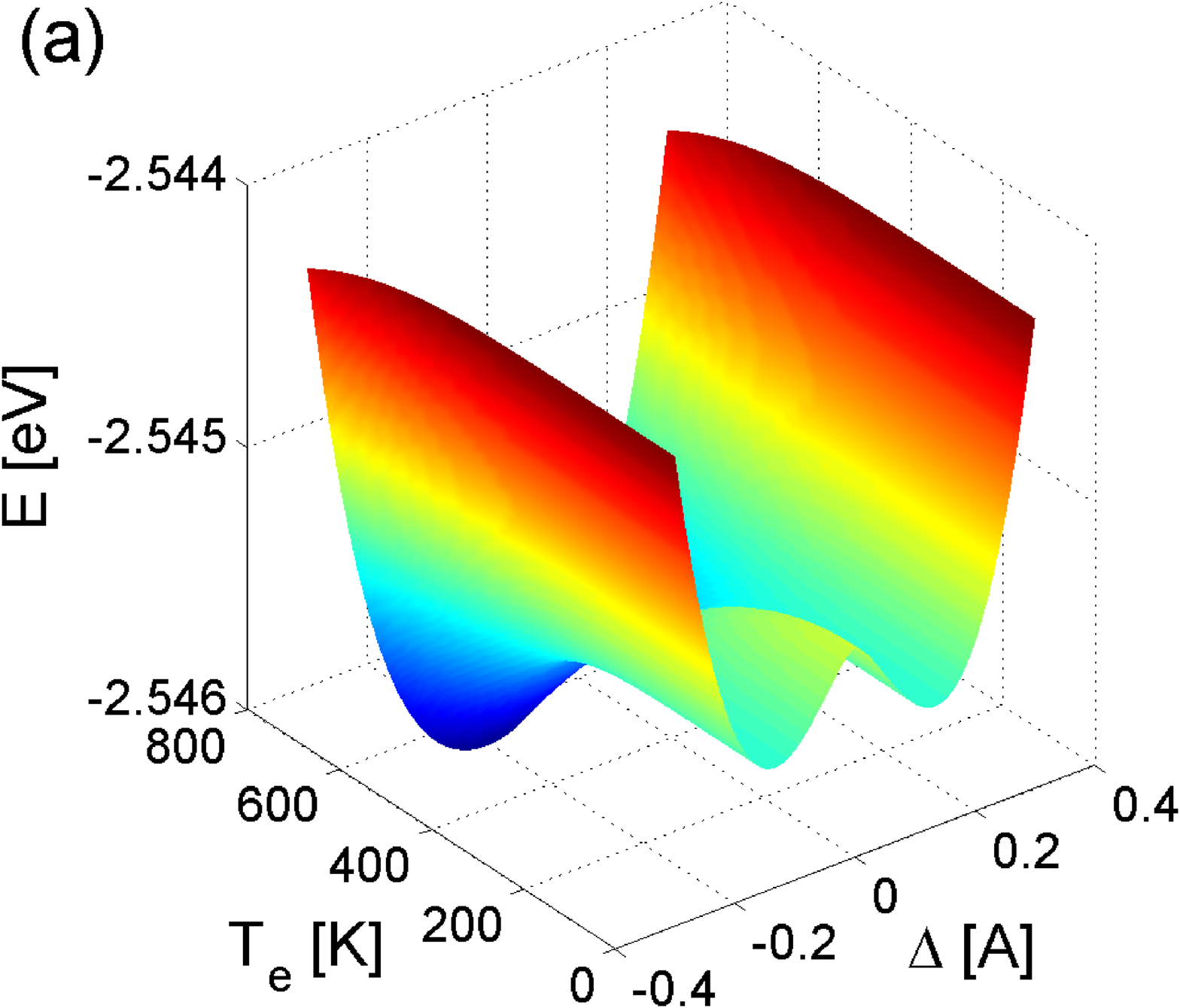}
\includegraphics[scale=0.25,clip]{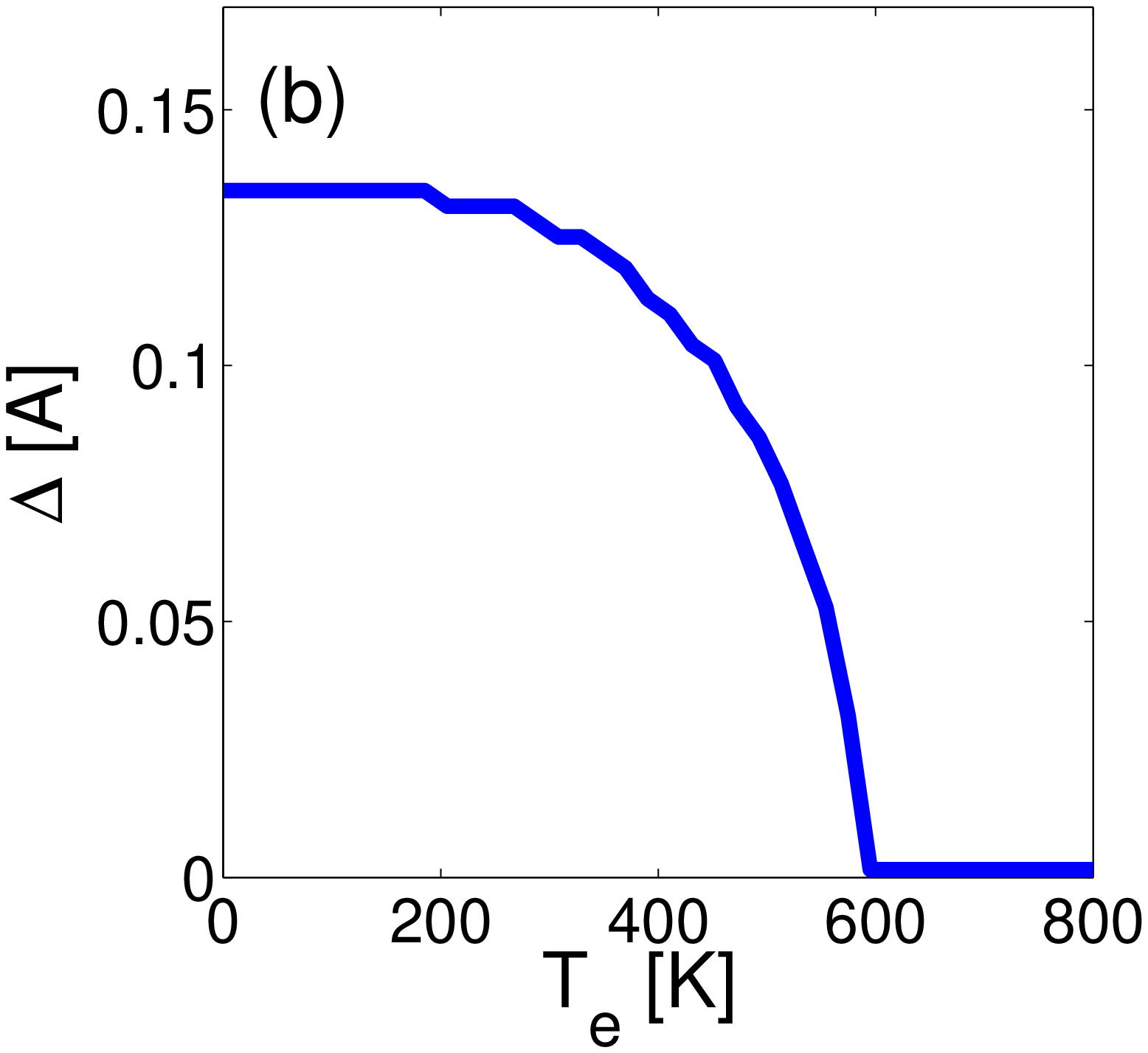}
\includegraphics[scale=0.25,clip]{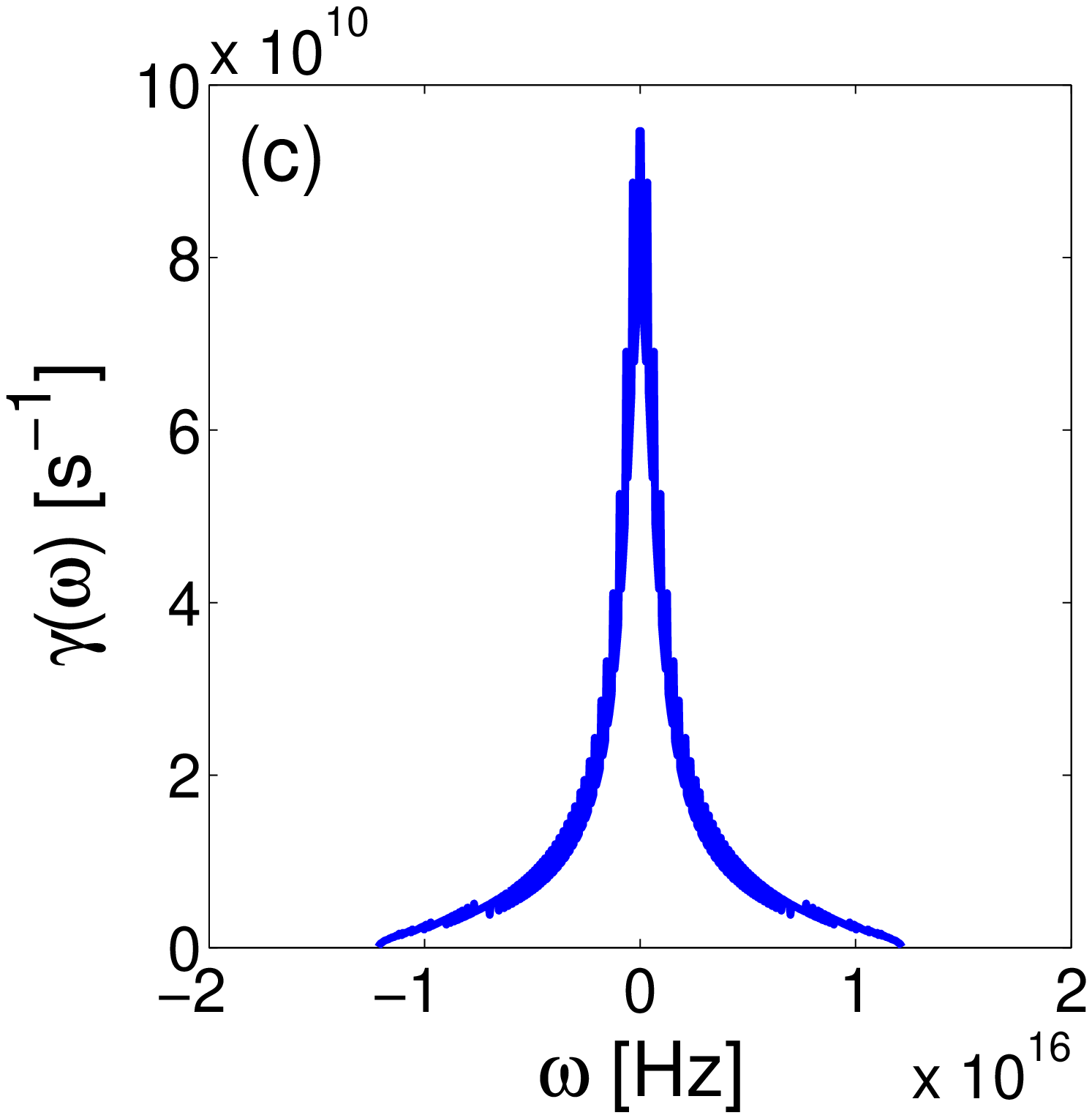}
\includegraphics[scale=0.25,clip]{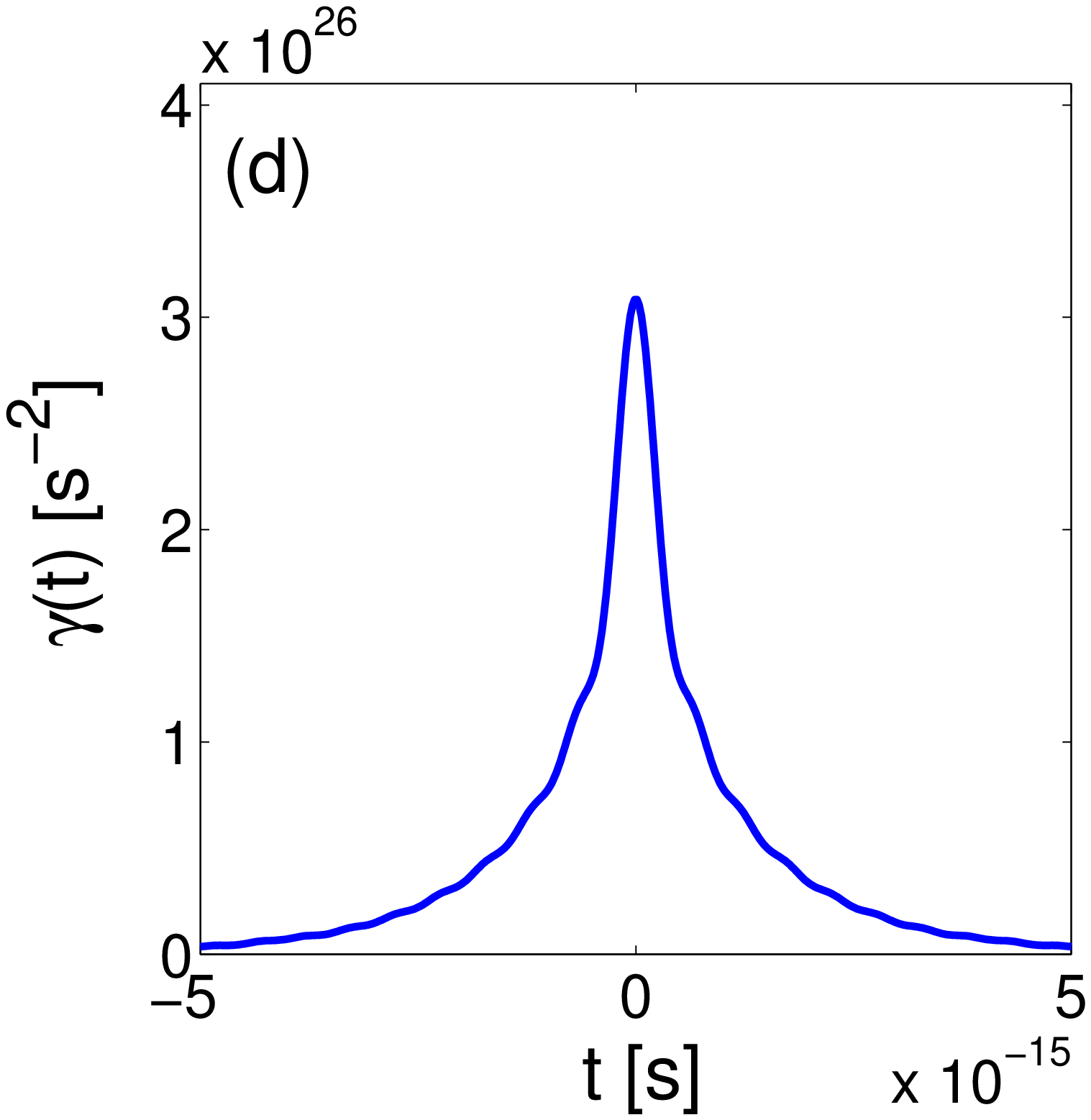}
\caption{(Color online)(a): the free energy per atom $\mathcal{E}$ as the function of electronic temperature $T_{e}$ and order parameter $\Delta$. (b): the order parameter $\Delta$ as a function of electronic temperature $T_{e}$, obtained by minimizing the free energy density $\mathcal{E}$. (c) and (d): the damping kernel $\gamma(t)$ and its Fourier transformation $\gamma(\omega)$ calculated for $\Delta=0$ and $T_{e}=1000$K. The basic model parameters for calculations here are given as $t_{0}=2$eV, $\alpha=0.167$eV/\AA, $M=1.35\times10^{-26}$kg, $K=1.2\text{kg}\cdot\text{s}^{-2}$, and $N=2000$. }\label{FEGap}
\end{figure}

\emph{Calculation Results} We perform some calculations based on the established theory above. We choose the following
parameters, $t_{0}=2$~eV, $\alpha=0.167$~eV/\AA, $M=1.35\times 10^{-26}$~kg, $K=1.2$~kg$\cdot\text{s}^{-2}$, which gives
the oscillation frequency $\Omega=6\pi$~THz. In the numerical calculation, we consider a system with $N=2000$ unit cells. The free energy $\mathcal{E}(\Delta,T_{e})$ in Eq.~(\ref{FEden}) calculated for
$\Delta\in[-0.3,0.3]\text{\AA}$ and $T_{e}\in[1,800]$K are shown in Fig.~\ref{FEGap}(a). At high temperature, the minimum of $\mathcal{E}$ is at $\Delta=0$ corresponding to the undistorted configuration (I); while at low temperature, $\mathcal{E}$ has the form of double well potential, and the distorted configuration (II) is stable. Fig.~\ref{FEGap}(b) shows the $T_{e}$ dependence of the stable order parameter $\Delta$($>0$), which is in fact the solution of the equation,
\begin{eqnarray}
4K=\frac{1}{N}\sum_{k}\frac{16\alpha^{2}\sin^{2}ka}{\Lambda_{k}}\text{tanh}\frac{\beta\Lambda_{k}}{2},\label{GapE}
\end{eqnarray}
obtained by setting $\mathsf{f}=0$ in Eq.~(\ref{FOCE}). The phase transition temperature is found to be  about $600$K for the chosen parameters here. The solutions thus recover the mean-field results for the CDW phase transition\cite{Gruner}.

With the given parameters, the damping kernel function $\gamma(t)$ and its Fourier transformation $\gamma(\omega)$ are also calculated with their expressions for $\Delta=0$ and $T_{e}=1000$K, as shown in Fig.~\ref{FEGap} (c) and (d). According to the qualitative estimation above, the cutoff frequency $\omega_{c}=1.2\times10^{16}$Hz, and $\gamma(\omega)\sim1.7\times10^{9}\text{s}^{-1}$; while the correlation time $t_{c}=5.2\times10^{-16}$s, and $\gamma(t)\sim 2\times10^{25}\text{s}^{-2}$. This indeed captures the qualitative features of the calculated damping kernel. Furthermore, the damping coefficient in Eq.~(\ref{equD1}) is calculated as $\Gamma=0.25\text{ps}^{-1}$ by integrating $\gamma(t)$ from $0$ to $5$fs. This parameter will be exploited in the following dynamics simulations.

\begin{figure}[htbp]
\includegraphics[scale=0.26,clip]{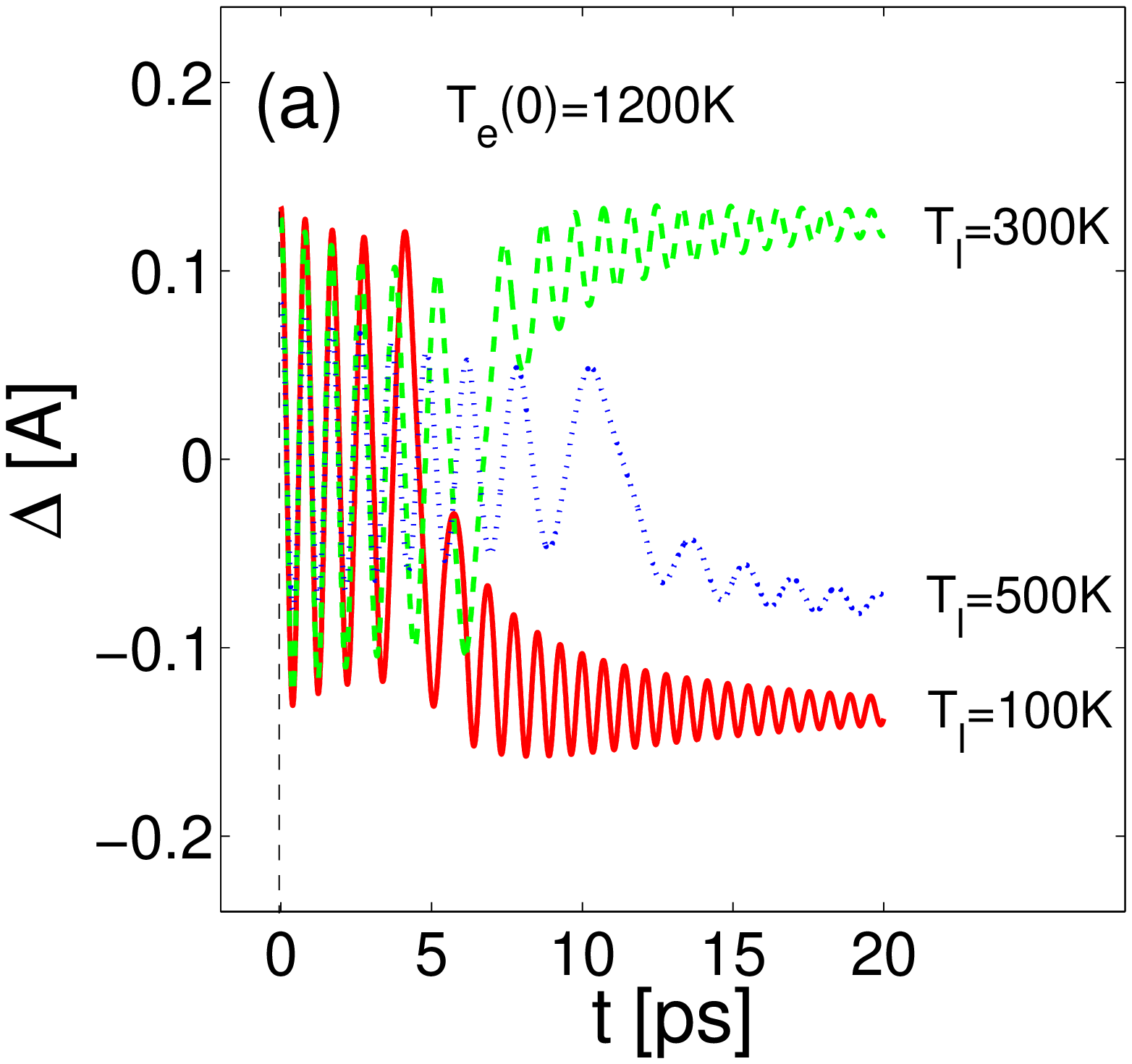}
\includegraphics[scale=0.26,clip]{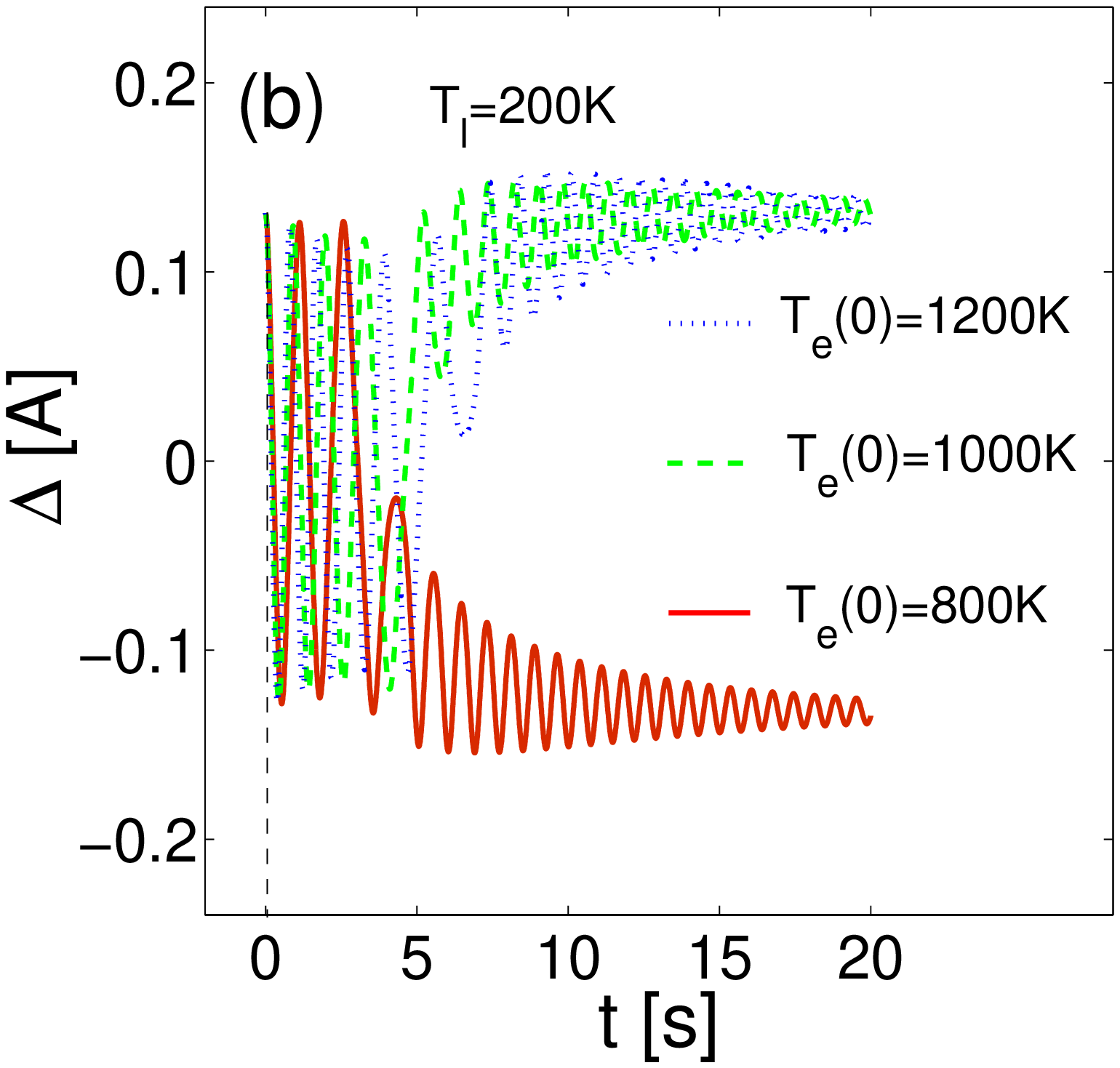}
\caption{(Color online)(a): order parameter dynamics for three different lattice temperatures $T_{l}=100,300,500$K with fixed initial electronic temperature $T_{e}(0)=1200$K. (b) order parameter dynamics for three different initial electronic temperatures $T_{e}(0)=800,1000,1200$K with fixed lattice temperature $T_{l}=200$K. The basic model parameters are the same as Fig.~\ref{FEGap}, and the damping coefficient $\Gamma=0.25\text{ps}^{-1}$, the electronic temperature decay time $\kappa=5$ps, and the discretized time step $\delta t=5$fs.}\label{Dym}
\end{figure}

Now we simulate the dynamics of order parameter $\Delta$ based on Eq.~(\ref{equD1}) and Eq.~(\ref{Tdcay}). The decay
time of $T_e$ is chosen to be $\kappa=5$ps.  In the calculation, we discretize the equations with $\delta t=5$fs between
two time steps. The initial order parameter $\Delta(0)$ is determined from Eq.~(\ref{GapE}) with given lattice
temperature $T_{l}$, and $\dot{\Delta}(0)=0$. The initial electronic temperature $T_{e}(0)$ depends on the fluence of
the pump laser, and is an input parameter in our simulations. The simulations with various $T_l$ and $T_e(0)$ are
performed.  The results with three different $T_{l}=100,300,500$K and fixed $T_{e}(0)=1200$K are depicted in
Fig.~\ref{Dym}(a), and the results with three different $T_{e}(0)=800,1000,1200$K and fixed $T_{l}=200$K are shown in Fig.~\ref{Dym}(b). Several features of the dynamics can be observed in the figures. First, the oscillation frequency of the order parameter is lower than the bare frequency $\Omega=6\pi$~THz of the phonon mode $(o,0)$, because of the coupling between the phonon and electrons. Second, the dynamics can be divided into two regions. When the electronic temperature $T_{e}$ is above the phase transition temperature, the amplitude of the oscillation is large while frequency
is low; when $T_{e}$ is below the phase transition temperature, the order parameter is trapped in one of the double wells, and it oscillates with smaller amplitude but higher frequency. This reflects the symmetry breaking when the electronic temperature crosses the phase transition temperature. Third, switching of the order parameter is possible in some cases as long as $T_{e}(0)$ is larger than the phase transition temperature. This suggests the possibility to control the Peierls systems optically in the pecosecond time scale for practical applications.

\emph{Conclusion} The photoexcited dynamics of order parameter in Peierls chain has been investigated within the microscopic quantum theory in the limit that the electron subsystem can be treated as quasi-equilibrium state with an exponentially decaying effective temperature. The optical phonon mode responsible for the Peierls distortion is coupled to the electron subsystem, and its dynamical equation is established with the help of the density matrix technique. The driven force acting on the order parameter and the damping coefficient are obtained in terms of the microscopic parameters in the Peierls model. We further apply the theory to simulate the recovery dynamics of the order parameter in the photoexcited Peierls chain. It is found that the oscillation frequency and amplitude will be significantly changed when the effective electronic temperature crosses the phase transition temperature, and switching of the order parameter may happen. Our theory describes the physical processes microscopically, and is not restricted to the temperature range near the phase transition as in the phenomenological Ginzburg-Landau theory. The theory can be further generalized to understand the photoexcited dynamics in more complicated Peierls-type systems and might have potential applications

\begin{acknowledgements}
This work was supported in part by the Hong Kong's University Grant Council via grant AoE/P-04/08. W. Q. Chen was partly
supported by NSFC project No. 11204186, and F. C. Zhang was partly supported by  NSFC project No. 11274269. 		 
\end{acknowledgements}


\begin{thebibliography}{99}
\bibitem{OrdP} P. M. Chaikin and T. C. Lubensky, \emph{Principles of
condensed matter physics}(Cambridge University Press, 1995).
\bibitem{MicroM} H. Kronm\"{u}ller and M. F\"{a}hnle, \emph{Micromagnetism and the Microstructure of Ferromagnetic Solids}(Cambridge University Press, 2003).
\bibitem{probe1} K. Kenji, M. Hase, H. Harima, S. Nakashima, M. Tani, K. Sakai, H. Negishi, and M. Inoue, Phys. Rev. B \textbf{58}, R7484 (1998).
\bibitem{probe2} J. Demsar, K. Biljakovi\'{c}, and D. Mihailovic, Phys. Rev. Lett. \textbf{83}, 800 (1999).
\bibitem{probe3} A. Tomeljak, H. Scha¨fer, D. St\"{a}dter, M. Beyer, K. Biljakovic, and J. Demsar, Phys. Rev. Lett. \textbf{102}, 066404 (2009).
\bibitem{probe4} R. Yusupov, T. Mertelj, V. V. Kabanov, S. Brazovskii, P. Kusar, J.-H. Chu, I. R. Fisher, and D. Mihailovic, Nat. Phys. \textbf{6}, 681 (2010).
\bibitem{ARPES1} L. Perfetti, P. A. Loukakos, M. Lisowski, U. Bovensiepen, H. Berger, S. Biermann,
P. S. Cornaglia, A. Georges, and M. Wolf, Phys. Rev. Lett. \textbf{97}, 067402 (2006).
\bibitem{ARPES2}F. Schmitt, P. S. Kirchmann, U. Bovensiepen, R. G. Moore, L. Rettig, M. Krenz,
J.-H. Chu, N. Ru, L. Perfetti, D. H. Lu, M. Wolf, I. R. Fisher, and Z.-X. Shen, Science,\textbf{321}, 1649 (2008).
\bibitem{ARPES3} F. Schmitt, P. S. Kirchmann, U. Bovensiepen, R. G. Moore, J.-H. Chu, D. H. Lu, L. Rettig, M. Wolf, I. R. Fisher, and Z.-X. Shen, New J. Phys. \textbf{13}, 063022 (2011).
\bibitem{ARPES4} J. C. Petersen, S. Kaiser, N. Dean, A. Simoncig, H.Y. Liu, A. L. Cavalieri, C. Cacho, I. C. E. Turcu, E. Springate, F. Frassetto, L. Poletto, S. S. Dhesi, H. Berger, and A. Cavalleri1, Phys. Rev. Lett. \textbf{107} 177402 (2011).
\bibitem{ARPES5} S. Hellmann, T. Rohwer, M. Kall\"{a}ne, K. Hanff, C. Sohrt, A. Stange, A. Carr, M.M. Murnane, H.C. Kapteyn, L. Kipp, M. Bauer, and K. Rossnagel, Nat. Comm. \textbf{3},1069 (2012).
\bibitem{ARPES6} H. Y. Liu, I. Gierz, J. C. Petersen, S. Kaiser, A. Simoncig, A. L. Cavalieri, C. Cacho, I. C. E. Turcu, E. Springate, F. Frassetto, L. Poletto, S. S. Dhesi, Z.-A. Xu, T. Cuk, R. Merlin, and A. Cavalleri, Phys. Rev. B \textbf{88}, 045104 (2013).
\bibitem{HHM} G. De Filippis, V. Cataudella, E. A. Nowadnick, T. P. Devereaux, A. S. Mishchenko, and N. Nagaosa, Phys. Rev. Lett. \textbf{109}, 176402 (2012).
\bibitem{SSH} W. P. Su, J. R. Schrieffer, and A. J. Heeger, Phys. Rev. Lett. \textbf{42}, 1698 (1979).
\bibitem{supp} See Supplemental Material for details.
\bibitem{denm} K. Blum, \emph{Density Matrix Theory and Applications} (Springer-Verlag, Berlin Heidelberg, 2012).
\bibitem{Gruner} G. Gr\"{u}ner, \emph{Density Waves in Solids} (Addison-Wesley, Reading, MA, 1994).
\end{thebibliography}
\end{document}